\documentclass[aps,prd,secnumarabic,twocolumn,nofootinbib,superscriptaddress,floatfix]{revtex4}

\usepackage{graphics}
\usepackage{color}
\usepackage{graphicx}
\usepackage{amssymb}
\usepackage{amsmath}
\usepackage{epstopdf}
\usepackage{xspace}

\usepackage{mathbbol}
\usepackage{wasysym} 
\usepackage{bbm}     
 
\usepackage{rotating}
\usepackage{dcolumn}
\usepackage{bm}
\usepackage{longtable}
\usepackage{multirow}
\usepackage{enumitem}

\usepackage{ifpdf}
\ifpdf
\usepackage[pdftex]{hyperref}
\else
\usepackage[hypertex]{hyperref}
\fi

\hypersetup{
 pdftitle={},%
 pdfauthor={},%
 pdfsubject={},%
 pdfkeywords={},%
 pdfstartview={},%
 bookmarksopen=true, breaklinks=true, debug=true,
 colorlinks=true, linkcolor=blue, citecolor=red, urlcolor=red,
 hyperfigures=true
}

\def\figwid{3.30in}

\def\APS{The American Physical Society}

\def\figPermX#1#2#3{From \cite{#1} with kind permission, copyright ({#2}) {#3}}
\def\figPerm#1#2#3#4{{#1} from \cite{#2} with kind permission, copyright ({#3}) {#4}}

\def\figPermXAPS#1#2{\figPermX{#1}{#2}{\APS}}

\def\AfigPerm#1#2#3{\figPerm{Adapted}{#1}{#2}{#3}}
\def\AfigPermAPS#1#2{\AfigPerm{#1}{#2}{\APS}}

\def\FThing#1#2{#1~\ref{#2}}

\def\Thing#1#2{#1.~\ref{#2}}

\def\Fig#1{\Thing{Fig}{#1}}

\def\Tab#1{\FThing{Table}{#1}}

\def\beq{\begin{equation}}
\def\eeq{\end{equation}}
\def\beqa{\begin{eqnarray}}
\def\eeqa{\end{eqnarray}}

\newcommand{\ie}   {{\it i.e.,}~}

\newcommand{\mev}  {~\ensuremath{{\mathrm{MeV}}     }}

\newcommand{\Dze}   {\ensuremath{ D^0 }}

\newcommand{\Dstn} {\ensuremath{ D^{*0} }}

\newcommand{\DDbar}      {\ensuremath{ D   \bar D     }}

\newcommand{\DDst}    {\ensuremath{ D   \bar D^* }}

\newcommand{\DstnDn}  {\ensuremath{\Dstn \bar{\Dze}}}

\newcommand{\lap}   {\ensuremath{\Lambda_c^+ }}
\newcommand{\lam}   {\ensuremath{\Lambda_c^- }}
\newcommand{\lala}  {\ensuremath{\lap\lam }}

\def\Ups{\ensuremath{\Upsilon}}

\def\Unx#1#2{\ensuremath{\Ups({#1}{#2})}}
\def\UnS#1{\ensuremath{\Unx{#1}{S}}}
\def\UoneD{\ensuremath{\Unx{1}{D}}}

\def\UoneDT{\ensuremath{\Unx{1}{^3D_2}}}

\newcommand{\jpsi}{\ensuremath{J/\psi}}

\newcommand{\psip}{\ensuremath{\psi(2S)}}

\usepackage{relsize}
\def\babar{\mbox{\slshape B\kern-0.1em{\smaller A}\kern-0.1em
    B\kern-0.1em{\smaller A\kern-0.2em R}}}
\newcommand{\DZero}{\rm D\O}

\def\dmhf{\Delta m_{\rm hf}}

\newcommand{\piz}{{\ensuremath{{\pi^0}}}}

\newcommand{\dipi}{{\ensuremath{{\pi^+\pi^-}}}}
\newcommand{\dimu}{{\ensuremath{{\mu^+\mu^-}}}}

\def\simg{{\ \lower-1.2pt\vbox{\hbox{\rlap{$>$}\lower6pt\vbox{\hbox{$\sim$}}}}\ }}

\newcommand{\etab}{\ensuremath{{\eta_b(1S)}}}
\newcommand{\etac}{\ensuremath{{\eta_c(1S)}}}
\newcommand{\etacp}{\ensuremath{{\eta_c(2S)}}}

\newcommand{\hsubc}{\ensuremath{h_c(1P)}}
\newcommand{\hsubb}{\ensuremath{h_b(1P)}}
\newcommand{\hsubbOne}{\ensuremath{h_b(1P)}}
\newcommand{\hsubbTwo}{\ensuremath{h_b(2P)}}

\begin{document}

\title{\LARGE Developments in heavy quarkonium spectroscopy}
\thispagestyle{empty}

\renewcommand{\thefootnote}{\fnsymbol{footnote}}

\author{S.~Eidelman}
\affiliation{Budker Institute of Nuclear Physics SB RAS, Novosibirsk 630090, Russia}
\affiliation{Novosibirsk State University, Novosibirsk 630090, Russia}

\author{B.~K.~Heltsley}
\affiliation{Cornell University, Ithaca, NY 14853, USA}

\author{J.~J.~Hern\`andez-Rey}
\affiliation{Instituto de F\'isica Corpuscular,
Universitat de Val\`encia--CSIC,
Edificio de Investigaci\'on de Paterna,
Apdo. 22085,
E-46071 Valencia, Spain}

\author{S.~Navas}
\affiliation{Departamento de F\'isica Te\'orica y del Cosmos \& CAFPE, 
Universidad de Granada, 18071 Granada, Spain}

\author{C.~Patrignani}
\affiliation{Dipartimento de Fisica e INFN, Universit\`a de
  Genova. I-16146 Genova, Italy}

\date{\today}

\begin{abstract}
We summarize recent developments in heavy quarkonium
spectroscopy, relying on previous review articles for the bulk of
material available prior to mid-2010. This note is intended as a
mini-review to appear in the 2012 Review of Particle Physics
published by the Particle Data Group.
\end{abstract}

\maketitle
 A golden age for heavy quarkonium physics dawned a decade
ago, initiated by the confluence of exciting advances in 
quantum chromodynamics (QCD) and an explosion of related experimental
activity. The subsequent broad spectrum of breakthroughs, surprises, 
and continuing puzzles had not been anticipated. 
In that period, the BESII program concluded only to give birth to BESIII;
the $B$-factories and CLEO-c flourished;
quarkonium production and polarization 
measurements at HERA and the Tevatron matured; 
and heavy-ion collisions at RHIC opened a window
on the deconfinement regime.
For an extensive presentation of the status of 
heavy quarkonium physics, the reader is
referred to several
reviews~\cite{Brambilla:2004wf,Eichten:2007qx,Eidelman:2008zzc,Godfrey:2008nc,Barnes:2009zza,Pakhlova:2010zz,Brambilla:2010cs},
the last of which covers developments through the middle of 2010,
and which supplies some tabular information and phrasing reproduced
here (with kind permission, copyright 2011, Springer).
This note focuses solely on experimental developments in
heavy quarkonium spectroscopy, and in particular on those
too recent to have been included in \cite{Brambilla:2010cs}.

\begin{table*}[t]
\caption{New {\it conventional} states in the $c\bar{c}$, $b\bar{c}$, and
$b\bar{b}$ regions, ordered by mass. Masses $m$ and widths $\Gamma$ represent
the weighted averages from the listed sources.
Quoted uncertainties reflect quadrature summation from individual experiments.
In the Process column, the decay mode of the new state 
claimed is indicated in parentheses.
Ellipses (...) indicate inclusively selected event topologies;
\ie additional particles not required by the Experiments to be present.
A question mark (?) indicates an unmeasured value.
For each Experiment a citation is given, as well as
the statistical significance in number of
standard deviations (\#$\sigma$), or ``(np)'' for ``not provided''. 
The Year column gives the date of first measurement cited.
The Status column indicates that the state 
has been observed 
by at most one ({\color{red} NC!}-needs confirmation) or
at least two independent experiments with significance of $>$5$\sigma$ (OK).
The state labelled $\chi_{c2}(2P)$ has previously been called $Z(3930)$.
See also the reviews in 
\cite{Brambilla:2004wf,Eichten:2007qx,Eidelman:2008zzc,Godfrey:2008nc,Barnes:2009zza,Pakhlova:2010zz,Brambilla:2010cs}.
Adapted from \cite{Brambilla:2010cs} with kind permission, copyright (2011), Springer.
} 

\setlength{\tabcolsep}{0.30pc}
\begin{center}
\begin{tabular}{lccclccc}
\hline\hline
\rule[10pt]{-1mm}{0mm}
 State & $m$~(MeV) & $\Gamma$~(MeV) & $J^{PC}$ & \ \ \ \ Process~(mode) & 
     Experiment~(\#$\sigma$) & Year & Status \\[0.7mm]
\hline
\rule[10pt]{-1mm}{0mm}
$\hsubc$ & $3525.41\pm 0.16$ & $<$1 & $1^{+-}$ &
    $\psi(2S)\to\pi^0\, (\gamma\etac)$ & 
    {\color{red} CLEO}~\cite{Rubin:2005px,Rosner:2005ry,Dobbs:2008ec}~(13.2) &
    2004 & OK \\ [0.7mm]
& & & &    $\psi(2S)\to\pi^0\, (\gamma ...)$ & 
    {\color{red} CLEO}~\cite{Rubin:2005px,Rosner:2005ry,Dobbs:2008ec}~(10), BES~\cite{Ablikim:2010rc}~(19) & & \\ [0.7mm]
& & & & $p\bar p\to (\gamma \eta_{c})\to (\gamma\gamma\gamma)$ &
    E835~\cite{Andreotti:2005vu}~(3.1)& &\\[0.7mm]
& & & &    $\psi(2S)\to\pi^0\, (...)$ & 
    BESIII~\cite{Ablikim:2010rc}~(9.5) & & \\ [1.79mm]
$\etacp$ & $3638.9\pm1.3$ &10$\pm$4 &$0^{-+}$ &
    $B\to K\, (K_S^0K^-\pi^+)$ &
     {\color{red} Belle}~\cite{Choi:2002na,Vinokurova:2011dy}~(6.0) & 2002 & OK\\[0.7mm]
& & & & $e^+e^-\to e^+e^-\,(K_S^0 K^-\pi^+)$ &
     \babar~\cite{delAmoSanchez:2011bt,Aubert:2003pt}~(7.8), &  & \\[0.7mm]
& & & & & CLEO~\cite{Asner:2003wv}~(6.5), 
           Belle~\cite{Nakazawa:2008zz}~(6) & &\\[0.7mm]
& & & & $e^+e^-\to J/\psi\, (...)$ &
     \babar~\cite{Aubert:2005tj}~(np), Belle~\cite{Abe:2007jn}~(8.1)  & &\\[1.79mm]
$\chi_{c2}(2P)$ & $3927.2\pm2.6$ & 24$\pm$6 & $2^{++}$ &
     $e^+e^-\to e^+e^- (D\bar{D})$ &
     {\color{red} Belle}~\cite{Uehara:2005qd}~(5.3), \babar~\cite{:2010hka,delAmoSanchez:2010jr}~(5.8) & 2005 & OK\\ [1.79mm]
$B_c^+$ & $6277\pm6$ & - & $0^-$ & $\bar{p}p\to (\pi^+\jpsi) ...$
        & {\color{red} CDF}~\cite{Abe:1998wi,Aaltonen:2007gv}~(8.0),
          \DZero~\cite{Abazov:2008kv}~(5.2) & 2007 & OK\\[1.79mm]
$\eta_b(1S)$ & $9395.8\pm3.0$  & 12.4$^{+12.7}_{-5.7}$ & 0$^{-+}$ &
     $\Unx{3}{S}\to\gamma\,(...)$ &
     {\color{red} \babar}~\cite{:2008vj}~(10),
     CLEO~\cite{Bonvicini:2009hs}~(4.0)  & 2008 & OK\\[0.7mm]
  & & & & $\Unx{2}{S}\to\gamma\,(...)$ &
     \babar~\cite{:2009pz}~(3.0)  & & \\[0.7mm]
  & & & & $\Unx{5}{S}\to\dipi\gamma\,(...)$ & Belle~\cite{Adachi:2011ch}~(14) &
     & \\[3.79mm]
\hsubbOne & $9898.6\pm1.4$ & ? & $1^{+-}$ &
$\UnS{5}\to\dipi\,(...)$ &
Belle~\cite{Adachi:2011ji,Adachi:2011ch}~(5.5) & 2011 & {\color{red} NC!} \\[0.7mm]
    &&&&$\UnS{3}\to\piz\,(...)$ & \babar~\cite{Lees:2011zp}~(3.0) & & 

\\[3.79mm]
\UoneDT\  & $ 10163.7\pm1.4$ & ? & 2$^{--}$ &
     $\UnS{3}\to\gamma\gamma\,(\gamma\gamma\UnS{1})$ & 
     CLEO~\cite{Bonvicini:2004yj}~(10.2) & 2004 & OK\\[0.7mm]
&&&&     $\UnS{3}\to\gamma\gamma\,(\pi^+\pi^-\UnS{1})$ & 
     \babar~\cite{Sanchez:2010kz}~(5.8) & & \\[0.7mm]
&&&&     $\UnS{5}\to\dipi\,(...)$ & Belle~\cite{Adachi:2011ji}~(2.4) & &\\[3.79mm]
\hsubbTwo & $10259.8^{+1.5}_{-1.2}$ & ? & $1^{+-}$ &
$\Upsilon(5S)\to\dipi\,(...)$ & Belle~\cite{Adachi:2011ji}~(11.2) & 2011 & {\color{red} NC!} \\[3.79mm]
$\chi_{bJ}(3P)$ & $10530\pm10$ & ? & ? & $pp\to (\gamma\mu^+\mu^-)...$ & ATLAS~\cite{Aad:2011ih}~($>$6) & 2011 &  {\color{red} NC!} \\[2.79mm]
\hline\hline
\end{tabular}
\end{center}
\label{tab:convent}
\end{table*}

\Tab{tab:convent} lists properties of newly observed
conventional heavy quarkonium states, where ``newly'' is
interpreted to mean within the past decade. The $h_c$ is the $^1P_1$ state of 
charmonium, singlet partner of the long-known $\chi_{cJ}$ triplet
$^3P_J$. The $\etacp$ is the first excited state of
the pseudoscalar ground state \etac, lying just
below the mass of its vector counterpart, $\psi(2S)$. 
The state originally dubbed $Z(3930)$ is now regarded
by many as the first observed $2P$ state of $\chi_{cJ}$,
the $\chi_{c2}(2P)$.
The first $B$-meson seen that
contains charm is the $B_c^+$. The ground state of
bottomonium is the $\eta_b(1S)$, recently confirmed
with a second observation of more than 5$\sigma$ significance. 
The \UoneD\  is the
lowest-lying $D$-wave triplet of the $b\bar{b}$ system.
Both the $\hsubbOne$, the bottomonium counterpart of $\hsubc$,
and the next excited state, $\hsubbTwo$, were very 
recently observed by Belle~\cite{Adachi:2011ji}, 
as described further below,
in dipion transitions from either the $\Upsilon(5S)$ or $Y_b(10888)$.
All fit into their respective spectroscopies roughly 
where expected. Their exact masses, production mechanisms, and decay modes
provide guidance to their descriptions within QCD.
The $h_b(nP)$ states still need experimental confirmation at
the 5$\sigma$ level, as does the $\chi_{bJ}(3P)$ triplet.

\begin{table*}[t]
\caption{As in \Tab{tab:convent}, but for new 
{\it unconventional} states in the $c\bar{c}$ and
$b\bar{b}$ regions, ordered by mass. 
For $X(3872)$, the values given are based only upon 
decays to $\pi^+\pi^- J/\psi$. 
$X(3945)$ and $Y(3940)$ have been subsumed under 
$X(3915)$ due to compatible properties. 
The state known as $Z(3930)$ appears as
the $\chi_{c2}(2P)$ in \Tab{tab:convent}.
In some cases experiment still allows
two $J^{PC}$ values, in which case both appear. 
See also the reviews in 
\cite{Brambilla:2004wf,Eichten:2007qx,Eidelman:2008zzc,Godfrey:2008nc,Barnes:2009zza,Pakhlova:2010zz,Brambilla:2010cs}.
Adapted from \cite{Brambilla:2010cs} with kind permission, copyright (2011), Springer.
 } 
\setlength{\tabcolsep}{0.21pc}
\label{tab:unconvent}
\begin{center}
\begin{tabular}{lccclccc}
\hline\hline
\rule[10pt]{-1mm}{0mm}
 State & $m$~(MeV) & $\Gamma$~(MeV) & $J^{PC}$ & \ \ \ \ Process~(mode) & 
     Experiment~(\#$\sigma$) & Year & Status \\[0.7mm]
\hline
\rule[10pt]{-1mm}{0mm}
$X(3872)$& 3871.68$\pm$0.17 & $<1.2$ &
    $1^{++}/2^{-+}$
    & $B\to K\, (\pi^+\pi^-J/\psi)$ &
    {\color{red} Belle} \cite{Choi:2003ue,Choi:2011fc}~(12.8), 
    \babar~\cite{Aubert:2008gu}~(8.6) & 2003 & OK \\[0.7mm]
& & & & $p\bar p\to (\pi^+\pi^- J/\psi)+ ...$ &
    CDF~\cite{Acosta:2003zx,Abulencia:2006ma,Aaltonen:2009vj}~(np), \DZero~\cite{Abazov:2004kp}~(5.2) & &\\[0.7mm]
& & &   & $B\to K\, (\omega J/\psi)$ &
    Belle~\cite{Abe:2005ix}~(4.3),
    \babar~\cite{delAmoSanchez:2010jr}~(4.0) & &\\[0.7mm]
& & & & $B\to K\, (\DstnDn)$ &
    Belle~\cite{Gokhroo:2006bt,Aushev:2008su}~(6.4), 
    \babar~\cite{Aubert:2007rva}~(4.9) & &\\[0.7mm]
& & & & $B\to K\, (\gamma J/\psi)$ &
    Belle~\cite{Bhardwaj:2011dj}~(4.0), \babar~\cite{Aubert:2006aj,Aubert:2008rn}~(3.6)&&\\[0.7mm]
& & & & $B\to K\, (\gamma \psi(2S))$ &
    \babar~\cite{Aubert:2008rn}~(3.5),
    Belle~\cite{Bhardwaj:2011dj}~(0.4) & & \\[0.7mm]
& & & &  $pp\to (\pi^+\pi^- J/\psi)+ ...$   & 
    LHCb~\cite{Aaij:2012lhcb}~(np) & & \\[1.89mm]
$X(3915)$ & $3917.4\pm2.7$ & 28$^{+10}_{-~9}$ & $0/2^{?+}$ &
    $B\to K\, (\omega \jpsi)$ &
    {\color{red} Belle}~\cite{Abe:2004zs}~(8.1),
    \babar~\cite{Aubert:2007vj}~(19) & 2004 & OK\\ [0.7mm]
     & & & & $e^+e^-\to e^+e^-\, (\omega\jpsi)$ &
    Belle~\cite{Uehara:2009tx}~(7.7),
    \babar~\cite{delAmoSanchez:2010jr}~(np) &&\\[1.89mm]
$X(3940)$ & $3942^{+9}_{-8}$ & $37^{+27}_{-17}$ & $?^{?+}$ &
     $e^+e^-\to J/\psi\,(\DDst)$ &
     {\color{red} Belle}~\cite{Abe:2007sya}~(6.0) & 2007 & {\color{red} NC!}\\ [0.7mm]
&&&& $e^+e^-\to J/\psi\, (...)$ &
     {\color{red} Belle}~\cite{Abe:2007jn}~(5.0) & \\ [1.89mm]
$G(3900)$ & $3943\pm21$ & 52$\pm$11 & $1^{--}$ &
     $e^+e^-\to\gamma\,  (\DDbar)$ &
     {\color{red} \babar}~\cite{Aubert:2006mi}~(np),
     Belle~\cite{Pakhlova:2008zza}~(np)
     & 2007 & OK \\ [1.89mm]
$Y(4008)$ & $4008^{+121}_{-\ 49}$ & 226$\pm$97 & $1^{--}$ &
     $e^+e^-\to\,\gamma  (\pi^+\pi^-J/\psi)$ &
     {\color{red} Belle}~\cite{Belle:2007sj}~(7.4)
     & 2007 & {\color{red} NC!} \\[1.89mm]
$Z_1(4050)^+$ & $4051^{+24}_{-43}$ & $82^{+51}_{-55}$ & ?&
     $ B\to K\, (\pi^+\chi_{c1}(1P))$ &
     {\color{red} Belle}~\cite{Mizuk:2008me}~(5.0),
     \babar~\cite{Lees:2011ik}~(1.1)
  & 2008 & {\color{red} NC!}\\[1.89mm]
$Y(4140)$ & $4143.4\pm3.0$  & $15^{+11}_{-\ 7}$ & $?^{?+}$ &
     $B\to K\, (\phi J/\psi)$ &
     {\color{red} CDF}~\cite{Aaltonen:2009tz,Aaltonen:2011at}~(5.0) & 2009 & {\color{red} NC!}\\[1.89mm]
$X(4160)$ & $4156^{+29}_{-25} $ & $139^{+113}_{-65}$ & $?^{?+}$ &
     $e^+e^-\to\jpsi\,(\DDst)$ &
     {\color{red} Belle}~\cite{Abe:2007sya}~(5.5) & 2007 & {\color{red} NC!}\\[1.89mm]
$Z_2(4250)^+$ & $4248^{+185}_{-\ 45}$ &
     177$^{+321}_{-\ 72}$ &?&
     $ B\to K\, (\pi^+\chi_{c1}(1P))$ &
     {\color{red} Belle}~\cite{Mizuk:2008me}~(5.0),
     \babar~\cite{Lees:2011ik}~(2.0)
     & 2008 &{\color{red} NC!}\\[1.89mm]
$Y(4260)$ & $4263^{+8}_{-9}$ & 95$\pm$14 & $1^{--}$ &
     $e^+e^-\to\gamma\,  (\pi^+\pi^- J/\psi)$ &
     {\color{red} \babar}~\cite{Aubert:2005rm,Aubert:2008ic}~(8.0) & 2005 & OK \\ [0.7mm]
     &&&&& CLEO~\cite{He:2006kg}~(5.4), Belle~\cite{Belle:2007sj}~(15) & & \\ [0.7mm]
& & & & $e^+e^-\to (\pi^+\pi^- J/\psi)$ & CLEO~\cite{Coan:2006rv}~(11)& &\\[0.7mm]
& & & & $e^+e^-\to (\pi^0\pi^0 J/\psi)$ & CLEO~\cite{Coan:2006rv}~(5.1) & &\\[1.89mm]
$Y(4274)$ & $4274.4^{+8.4}_{-6.7}$ & $32^{+22}_{-15}$ & $?^{?+}$ &
     $B\to K\, (\phi J/\psi)$ &
     {\color{red} CDF}~\cite{Aaltonen:2011at}~(3.1) & 2010 & {\color{red} NC!}\\[1.89mm]
$X(4350)$ & $4350.6^{+4.6}_{-5.1}$ & $13.3^{+18.4}_{-10.0}$ & 0/2$^{++}$ &
     $e^+e^-\to e^+e^- \,(\phi\jpsi)$ &
     {\color{red} Belle}~\cite{Shen:2009vs}~(3.2) & 2009 & {\color{red} NC!}\\ [1.89mm]
$Y(4360)$ & $4361\pm13$ & 74$\pm$18 & $1^{--}$ &
     $e^+e^-\to\gamma\,  (\pi^+\pi^- \psip)$ &
     {\color{red} \babar}~\cite{Aubert:2006ge}~(np),
     Belle~\cite{:2007ea}~(8.0) & 2007 &  OK \\ [1.89mm]
$Z(4430)^+$ & $4443^{+24}_{-18}$ & $107^{+113}_{-\ 71}$ & ?&
     $B\to K\, (\pi^+\psi(2S))$ &
     {\color{red} Belle}~\cite{Choi:2007wga,Mizuk:2009da}~(6.4),
     \babar~\cite{:2008nk}~(2.4)
     & 2007 & {\color{red} NC!}\\[1.89mm]
$X(4630)$ & $4634^{+\ 9}_{-11}$ & $92^{+41}_{-32}$ & $1^{--}$ &
     $e^+e^-\to\gamma\, (\lala)$ &
     {\color{red} Belle}~\cite{Pakhlova:2008vn}~(8.2)  & 2007 & {\color{red} NC!}\\ [1.89mm]
$Y(4660)$ & 4664$\pm$12 & 48$\pm$15 & $1^{--}$ &
     $e^+e^-\to\gamma\, (\pi^+\pi^- \psi(2S))$ &
     {\color{red} Belle}~\cite{:2007ea}~(5.8)  & 2007 & {\color{red} NC!}\\ [3.89mm]
$Z_{b}(10610)^+$ & 10607.2$\pm$2.0 & 18.4$\pm$2.4 & $1^+$ &
       $\Upsilon(5S)\to\pi^-(\pi^+\,[b\bar{b}]\,)$ &
      {\color{red} Belle}~\cite{Adachi:2011XXX,Bondar:2011pd}~(16) & 2011 & {\color{red} NC!}\\[1.89mm]
$Z_{b}(10650)^+$ & 10652.2$\pm$1.5 & 11.5$\pm$2.2 & $1^+$ &
       $\Upsilon(5S)\to\pi^-(\pi^+\,[b\bar{b}]\,)$ &
    {\color{red} Belle}~\cite{Adachi:2011XXX,Bondar:2011pd}~(16)& 2011
    & {\color{red} NC!}\\[3.89mm]
$Y_b(10888)$ & 10888.4$\pm$3.0 & 30.7$^{+8.9}_{-7.7}$ & $1^{--}$ &
      $e^+e^-\to(\pi^+\pi^- \Upsilon(nS))$ &
      {\color{red} Belle}~\cite{Chen:2008pu,Abe:2007tk}~(2.0)& 2010 & {\color{red} NC!}\\[2.79mm]
\hline\hline
\end{tabular}
\end{center}
\end{table*}

Correspondingly, the menagerie of new,
heavy-quarkonium-like {\it unanticipated} 
states\footnote{For consistency with the literature, 
we preserve the use of $X$, $Y$, $Z$, and $G$, contrary
to the practice of the PDG, which exclusively
uses $X$ for unidentified states.}
is shown in \Tab{tab:unconvent}; notice that
just a handful have been experimentally confirmed.
None can unambiguously be assigned a place in
the hierarchy of charmonia or bottomonia; neither do 
any have a universally accepted unconventional
origin. The $X(3872)$ occupies a unique niche among the
unexplained states as both the first and the most intriguing. 
It is, by now, widely studied, yet its interpretation 
demands much more experimental attention. 
The $Y(4260)$ and $Y(4360)$ are vector states
decaying to $\dipi\jpsi$ and $\dipi\psip$, respectively,
yet, unlike most conventional vector charmonia, 
do not correspond to enhancements in the $e^+e^-$ hadronic
cross section. The three $Z_c^+$ and two $Z_b^+$ 
states, each decaying to a charged pion and conventional 
heavy quarkonium state, would be manifestly exotic, but
remain unconfirmed. 
Final states of the type $\Upsilon(nS)\dipi$ from $e^+e^-$ 
collisions acquired near the $\Upsilon(5S)$ have a 
lineshape differing somewhat from that of multi-hadronic events,
which suggested a new state
$Y_b(10888)$, distinct from $\Upsilon(5S)$, which could
be analogous to $Y(4260)$.
The nature of $Y_b(10888)$, if it does mimic the
behavior of the charmonium-region $Y$'s, could 
help to explain the observed (and otherwise unexpected) high rate of 
dipion transitions to $\Upsilon(nS)$ and $h_b(nP)$ seen in $e^+e^-$ 
collisions near the $\Upsilon(5S)$ region.
It could also provide insight into the $Z_b^+$ states,
which appear to be intermediate resonances in the dipion transitions.

\babar~\cite{:2008nk,Lees:2011ik} 
has searched for the three $Z_c^\pm$ states in the charmonium
mass region seen by Belle, and failed to observe any significant signals.
The approach taken in searching for $B\to Z^\pm K\to (c\bar{c}) K\pi$,
where $(c\bar{c})$ is $\psi(2S)$ or $\chi_{c1}$, is to first fit the
data for all reasonable $K\pi$ mass or angular structure, having 
demonstrated that the presence of one or more $Z$'s cannot be 
accommodated by this procedure. After doing so, the finding is that
some of what might be the Belle excess of events above Belle background gets
absorbed into the $K\pi$ structure of the \babar\ background. 
As shown in \Tab{tab:unconvent}, where Belle observes
signals of significances $5.0\sigma$, $5.0\sigma$, and $6.4\sigma$
for $Z_1(4050)^+$, $Z_2(4250)^+$, and $Z(4430)^+$, respectively,
\babar\ reports $1.1\sigma$, $2.0\sigma$, and $2.4\sigma$ effects,
setting upper limits on product branching fractions
that are not inconsistent with Belle's measured rates,
leaving the situation unresolved.

Although $\etacp$ measurements began to converge on a mass and
width nearly a decade ago, refinements are still in progress.
In particular, Belle~\cite{Vinokurova:2011dy} has revisited its
analysis of $B\to K\etacp$, $\etacp\to KK\pi$ decays with more
data and methods that account for interference between
the above decay chain, an equivalent one with the $\etac$ instead,
and one with no intermediate resonance. The net effect of
this interference is far from trivial; it shifts the apparent mass 
by $\sim$+10\mev\ and blows up the apparent width by a factor of six.
The updated $\etacp$ mass and width are in better accordance 
with other measurements than the previous treatment~\cite{Choi:2002na} 
not including interference. Complementing this measurement in
$B$-decay, \babar~\cite{delAmoSanchez:2011bt} updated their
previous~\cite{Aubert:2003pt}  $\etacp$ mass and width measurements
in two-photon production, where interference effects, judging from
studies of $\etac$, appear to be small. In combination, precision
on the $\etacp$ mass has improved dramatically.

New results on $\eta_b$, $h_b$, and $Z_b^+$ mostly come from Belle,
all from analyses of 121.4~fb$^{-1}$ of $e^+e^-$
collision data collected near the peak of the $\UnS{5}$ resonance.
They also appear in the same types of decay chains: 
$\UnS{5}\to\pi^- Z_b^+$, $Z_b^+\to \pi^+ (b\bar{b})$,
and, when the $b\bar{b}$ forms an $\hsubbOne$, frequently 
$\hsubbOne\to\gamma\eta_b$. 
Previous unsuccessful searches for $h_b$ focused on what was considered
the most easily detected production mechanism, $\UnS{3}\to\piz\hsubbOne$. 
In early 2011 \babar\ presented marginal evidence
for this transition at the $3\sigma$ level, at a mass near that
expected for zero hyperfine splitting.

\begin{figure}[t]
   \includegraphics*[width=\figwid]{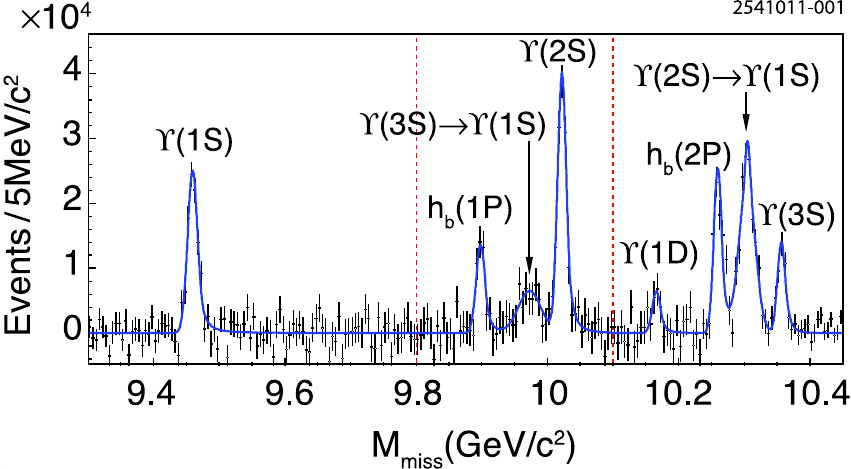}
   \caption{From Belle~\cite{Adachi:2011ji}, 
            the mass recoiling against $\dipi$ pairs, $M_{\rm miss}$,
            in $e^+e^-$ collision data 
            taken near the peak
            of the $\UnS{5}$ ({\it points with error bars}). 
            The smooth combinatoric and $K^0_S\to\dipi$ background 
            contributions have already been subtracted.
            The fit to the various labeled signal contributions 
            overlaid ({\it curve}).
            \AfigPermAPS{Adachi:2011ji}{2011}.
            }
\label{fig:dipirecoil}
\end{figure}

\begin{figure}[b]
   \includegraphics*[width=\figwid]{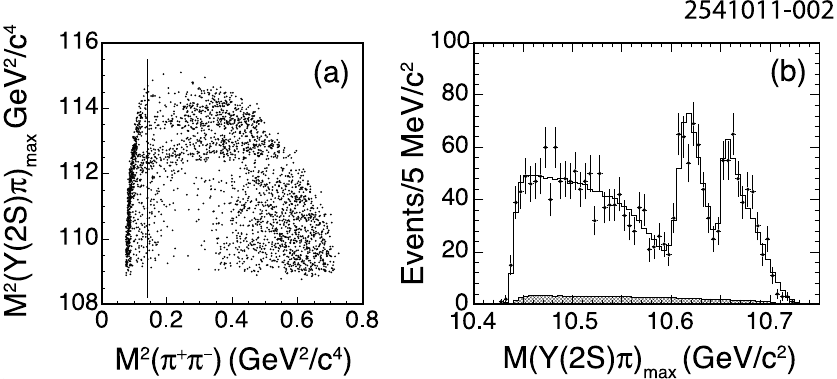}
   \caption{From Belle~\cite{Bondar:2011pd} $e^+e^-$ collision data 
            taken near the peak of the $\UnS{5}$
            for events with a $\dipi$-missing mass consistent
            with a $\UnS{2}$, (a) the maximum of the two possible single 
            $\pi^\pm$-missing-mass-squared combinations vs.~the
            $\dipi$-mass-squared; and (b) projection of the maximum of the two 
            possible single $\pi^\pm$-missing-mass combinations 
            ({\it points with error bars}) overlaid with a fit
            ({\it curve}). Events to the left of the
            vertical line in (a) are excluded from further analysis.
            The two horizontal stripes in (a) and two peaks in (b) 
            correspond to the two $Z_b^+$ states.
            \AfigPermAPS{Bondar:2011pd}{2011}.
            }
\label{fig:scat}
\end{figure}

The Belle $h_b$ discovery analysis~\cite{Adachi:2011ji} selects hadronic
events and looks for peaks in the mass recoiling against $\dipi$ pairs,
the spectrum for which, after subtraction of smooth combinatoric
and $K_S^0\to\dipi$ backgrounds, appears in \Fig{fig:dipirecoil}.
Prominent and unmistakable $\hsubbOne$ and $\hsubbTwo$ peaks are present. 
This search was directly inspired by a new CLEO result~\cite{:2011uqa}, 
which found the surprisingly copious transitions
$\psi(4160)\to\dipi\hsubc$ and an indication that 
$Y(4260)\to\dipi\hsubc$ occurs at a comparable 
rate as the signature mode, $Y(4260)\to\dipi\jpsi$. The presence of 
$\UnS{n}$ peaks in \Fig{fig:dipirecoil} at rates two orders of magnitude
larger than expected for transitions requiring a heavy-quark spin-flip, 
along with separate studies with exclusive decays $\UnS{n}\to\dimu$,
allow precise calibration of the $\dipi$ recoil mass spectrum and
very accurate measurements of $\hsubbOne$ and $\hsubbTwo$ masses.
Both corresponding hyperfine splittings are consistent with zero within
an uncertainty of about 1.5\mev\ (lowered to $\pm1.1\mev$ for
$\hsubbOne$ in \cite{Adachi:2011ch}).

\begin{figure}[b]
   \includegraphics*[width=\figwid]{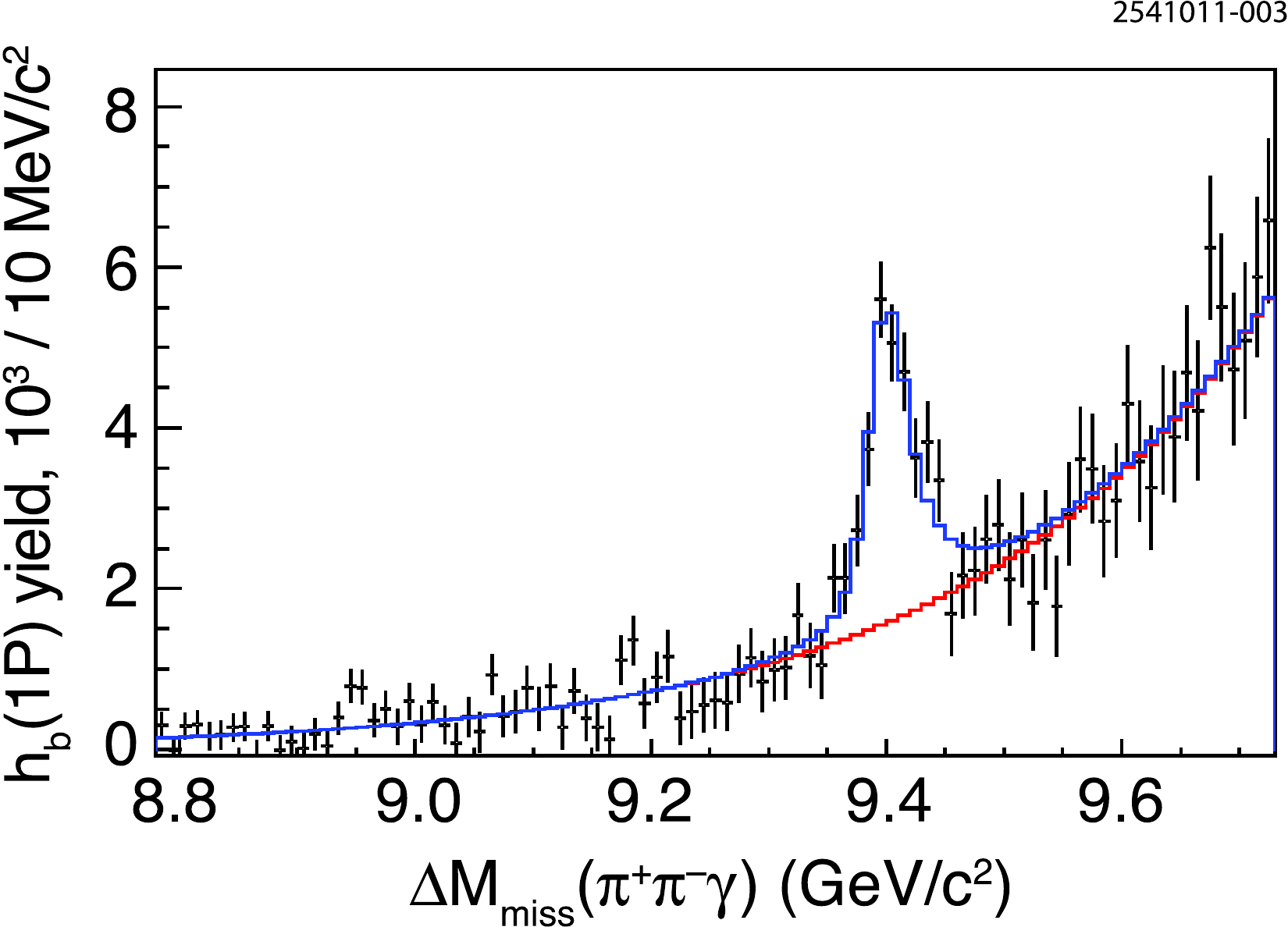}
   \caption{From Belle~\cite{Adachi:2011ch} $e^+e^-$ collision data 
            taken near the peak of the $\UnS{5}$,
            the $\hsubbOne$ event yield vs.~the mass recoiling against the
            $\dipi\gamma$ (corrected for misreconstructed $\dipi$), 
            where the $\hsubbOne$ yield is obtained
            by fitting the mass recoiling against the $\dipi$
            ({\it points with error bars}). The fit results 
            ({\it solid histograms}) for signal plus background
            and background alone are superimposed.
            \AfigPermAPS{Adachi:2011ch}{2011}.
            }
\label{fig:etab}
\end{figure}

Belle soon noticed that, for events in the peaks of 
\Fig{fig:dipirecoil}, there seemed to be two intermediate
charged states nearby. For example, \Fig{fig:scat} shows a Dalitz plot
for events restricted to the $\UnS{2}$ region of $\dipi$ recoil mass.
The two bands observed in the maximum of the two
$M[\pi^\pm\UnS{2}]^2$ values also appear for
$\UnS{1}$, $\UnS{3}$, $\hsubbOne$, and $\hsubbTwo$ samples, but 
do not appear in the respective $[b\bar{b}]$ sidebands. Belle fits
all subsamples to resonant plus non-resonant amplitudes, allowing
for interference (notably, between $\pi^-Z_b^+$ and $\pi^+Z_b^-$),
and finds consistent pairs of $Z_b^+$ masses for all bottomonium
transitions, and comparable strengths of the two states. 
Angular analysis favors a $J^P=1^+$ assignment for both $Z_b^+$
states, which must also have negative $G$-parity.
Transitions through
$Z_b^+$ to the $h_b(nP)$ saturate the observed $\dipi h_b(nP)$ cross sections.
The two masses of $Z_b^+$ states are just a few MeV above
the $B^*\bar{B}$ and $B^*\bar{B^*}$ thresholds, respectively.
The $Z_b^+$ cannot be simple mesons because
they are charged and have $b\bar{b}$ content.

The third Belle result to flow from these data is
confirmation of the $\etab$ and measurement of the
$\hsubbOne\to\gamma\etab$ branching fraction, expected
to be several tens of percent. To accomplish this,
events with the $\dipi$ recoil mass in the $\hsubbOne$
mass window and a radiative photon candidate are selected,
and the $\dipi\gamma$ recoil mass queried for correlation
with non-zero $\hsubb$ population in the $\dipi$ missing mass spectrun, 
as shown in \Fig{fig:etab}. A clear peak
is observed, corresponding to the $\etab$. A fit is performed
to extract the $\etab$ mass, and first measurements of its width and 
the branching fraction for $\hsubbOne\to\gamma\etab$
(the latter of which is $(49.8\pm6.8^{+10.9}_{-5.2})\%$).
The mass determination has comparable uncertainty to
and a larger central value (by 10\mev, or 2.4$\sigma$) than the
average of previous measurements, thereby reducing
the new world average hyperfine splitting by nearly 5\mev, 
as shown in \Tab{tab:etab}.

\begin{table}[t]
    \caption{Measured \etab\ masses and hyperfine splittings, by
    experiment and production mechanism.}
\label{tab:etab}
\setlength{\tabcolsep}{0.20pc}
\begin{center}
\begin{tabular}{cccc} 
\hline\hline
\rule[10pt]{-1mm}{0mm}
$m(\eta_b)$     & $\dmhf$  & Process & Ref. \\[0.7mm]
&&&($\chi^2$/d.o.f.)\\[0.7mm]
\hline
\rule[10pt]{-1mm}{0mm}
9394.2$^{+4.8}_{-4.9}$$\pm$2.0 &  $66.1^{+4.9}_{-4.8}$$\pm$2.0   &$\UnS{2}\to\gamma\eta_b$ & \babar~\cite{:2009pz} \\[1.5mm]
9388.9$^{+3.1}_{-2.3}$$\pm$2.7 & 71.4$^{+2.3}_{-3.1}$$\pm$2.7 &$\UnS{3}\to\gamma\eta_b$ & \babar~\cite{:2008vj} \\[1.5mm]
9391.8$\pm$6.6$\pm$2.0& 68.5$\pm$6.6$\pm$2.0  &$\UnS{3}\to\gamma\eta_b$ & CLEO~\cite{Bonvicini:2009hs}   \\[1.5mm]
$9391.0\pm2.8$ & $69.3\pm2.9$ & Above~\cite{Brambilla:2010cs} &Avg\footnote{An 
                 inverse-square-error-weighted average of 
                 the individual measurements appearing above, for
                 which all statistical and systematic errors
                 were combined in quadrature without accounting
                 for any possible correlations between them.
                 The uncertainty on this average is inflated
                 by the multiplicative factor $S$ 
                 if $S^2\equiv\chi^2$/d.o.f.$>$1}~(0.6/2)\\[1.5mm]
9401.0$\pm1.9^{+1.4}_{-2.4}$ & 59.3$\pm$1.9$^{+2.4}_{-1.4}$ & $\hsubbOne\to\gamma\eta_b$& Belle~\cite{Adachi:2011ch}\\[1.5mm]
$9395.8\pm3.0$& $64.5\pm3.0$ & All & Avg$^a$~(6.1/3)\\[0.7mm]
\hline
\hline
\end{tabular}
\end{center}
\end{table}

\begin{figure}[b]
   \includegraphics*[width=\figwid]{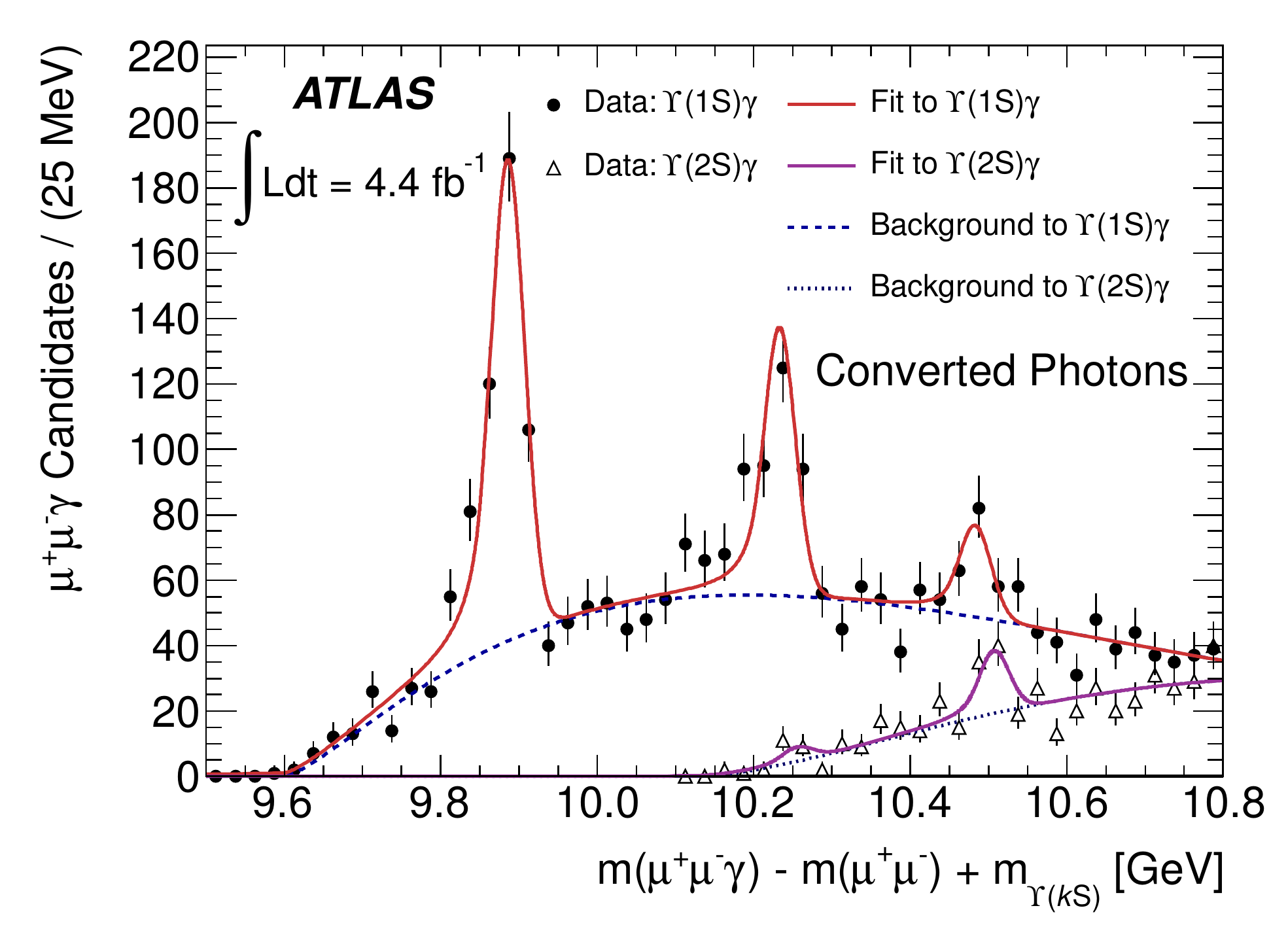}
   \caption{From ATLAS~\cite{Aad:2011ih} $pp$ collision data 
     ({\sl points with error bars}) 
            taken at $\sqrt{s}=7$~TeV, the effective mass
	    of $\chi_{bJ}(1P,2P,3P)\to\gamma\Upsilon(1S,2S)$ candidates
	    in which $\Upsilon(1S,2S)\to\mu^+\mu^-$
	    and the photon is reconstructed as an $e^+e^-$
	    conversion in the tracking system.
	    Fits ({\it smooth curves}) show significant 
            signals for each triplet (merged-$J$) 
	    on top of a smooth background.
            \figPermXAPS{Aad:2011ih}{2012}.
            }
\label{fig:atlas}
\end{figure}

The $\chi_{bJ}(nP)$ states have recently been observed
at the LHC by ATLAS~\cite{Aad:2011ih} for $n=1,2,3$, 
although in each case the three $J$ states are not distinguished
from one another. Events are sought which have both a photon
and an $\Upsilon(1S,2S)\to\mu^+\mu^-$ candidate which together
form a mass in the $\chi_b$ region. Observation of all three
$J$-merged peaks is seen at significance in excess of $6\sigma$
for both unconverted and converted photons. The mass plot for
converted photons, which provide better mass resolution, is 
shown in \Fig{fig:atlas}. This marks the first observation of
the $\chi_{bJ}(3P)$ triplet, quite near the expected mass.

\end{document}